\newcommand{\bea}{\begin{eqnarray}}
\newcommand{\eea}{\end{eqnarray}}
\newcommand{\bear}{\begin{eqnarray*}}
\newcommand{\eear}{\end{eqnarray*}}

\documentstyle[aps,preprint]{revtex}
\begin{document}

\draft

\title
{Integrable model of interacting XX and Fateev-Zamolodchikov  chains }

\author{
F. C. Alcaraz$^1$ and R. Z. Bariev$^{1,2}$}

\address{$^1$Departamento de F\'{\i}sica, 
Universidade Federal de S\~ao Carlos, 13565-905, S\~ao Carlos, SP
Brazil}

\address{$^2$The Kazan Physico-Technical Institute of the 
Russian Academy of Sciences, 
Kazan 420029, Russia}

\maketitle

\begin{abstract}
We consider the exact solution of a model of correlated particles,
which is presented as a system of interacting XX and Fateev-Zamolodchikov
chains. This model can also be considered  as a generalization of the multiband 
anisotropic $t-J$ model in  the case we restrict the 
site occupations to at
most two electrons. The exact solution is obtained for the eigenvalues 
and eigenvectors using the Bethe-ansatz method.
\end{abstract}

\pacs{PACS numbers: 75.10.Lp, 74.20-z, 71.28+d}

\narrowtext 

It is well known that in the limit of a strong Coulomb repulsion $U$ the 
traditional Hubbard model reduces to the so-called $t-J$ model [1]. The 
strong on-site correlations limit the site occupations to at most one 
electron. States with double occupation on a given site are 
energetically unfavorable
and can be projected out from the Hilbert space. 
In one dimension this model is integrable at the 
supersymmetrical point [2 - 5]. The exact solution can be obtained also for the 
extension of the supersymmetric spin-1/2 $t-J$ model (i.e. N = 2) to the 
case of  
arbitrary number $N$ of bands, having  a  $SU(N)$ symmetry [6,7], as well 
as for 
 their anisotropic version [8,9]. In some sense all these models  
can be considered as a system
of interacting XX and XXZ chains with size occupations limited to at most
one electron per site [2].

In this paper we consider a generalization of the multiband $t-J$ model 
for the case where the on-site correlations 
are enough strong to limit the site occupations 
to at most two electrons.  The exact integrable model we will present 
can be considered   as  a  
 system of interacting XX and spin-1 Fateev-Zamolodchikov chains [10,11].

In order to present our model let us consider initially a general model with 
$N$ distinct type of bands $(\alpha=1,\ldots,N)$ split into two disjoint 
groups $\aleph_1$ and $\aleph_2$.  Taking into account the constraint of 
maximum double occupancy in a given site we can have at most one 
electron in bands $\alpha \in \aleph_1$ and at most two electrons in bands 
$\alpha \in \aleph_2$. The states of a given site are denoted by 
$|\alpha,\beta>$ $(\alpha \leq \beta, \alpha,\beta=0,1,\ldots,N)$, where 
in the case $\alpha, \beta = 1, 2, \ldots N $ these are  
the bands where the electrons are 
located while  $\alpha =0$ or $\beta = 0$ denotes the absence of  electrons. 
As a conequence of this notation the state $|\alpha, \alpha>$ is forbidden if 
$\alpha \in \aleph_1$ and it is allowed if $\alpha \in \aleph_2$ or 
$\alpha =0$.
The most general Hamiltonian with nearest-neighbours interactions, 
in a lattice with L sites and periodic boundary condition, 
that conserves separately the electrons  in 
each band can be written as 
\bea
  H  &=& -\sum_{j=1}^L H_{j,j+1}  \nonumber\\  
H_{j,j+1} &=& 
\sum_{[\alpha],[\alpha']=0}^NW_{\alpha_1',\alpha_2',\alpha_3',\alpha_4'}
^{\alpha_1,\alpha_2,\alpha_3,\alpha_4}E_j^{\alpha_1,\alpha_2|\alpha_1',
\alpha_2'}E_{j+1}^{\alpha_3,\alpha_4|\alpha_3',\alpha_4'}
\eea
where $[\alpha]\equiv [\alpha_1',\alpha_2',\alpha_3',\alpha_4']$ is a 
permutation of 
 $[\alpha_1,\alpha_2,\alpha_3,\alpha_4]$  which keeps the order  
\bea
 \alpha_1\le\alpha_2, \;\;\; \alpha_1'\le\alpha_2',\;\;\; \alpha_3\le\alpha_4,
\;\;\;\alpha_3'\le\alpha_4'  ,\;\;\;
{\alpha_i} = 0,1,2,...,N. \nonumber 
\eea
The matrix 
$E^{\alpha,\beta|\alpha',\beta'}$ has all the elements zero except the 
element in the $\alpha,\beta$ line and $\alpha',\beta'$ column, with unit 
value, i. e. 
\bea
 E^{\alpha,\beta|\alpha',\beta'} = |\alpha\beta><\alpha'\beta'|.
\eea
We consider Hermitean Hamiltonians,  
\bea
W_{\alpha',\beta',\gamma',\delta'}^{\alpha \;,\beta \;,\gamma \;,\delta}&=& 
W_{\alpha \;\beta \;\gamma \;\delta}^{\alpha'\beta'\gamma'\delta'}, 
\eea
in the absence of  external fields, 
\bea
W_{0, \alpha, 0, 0}^{0, \alpha, 0, 0} = W_{0, 0, 0, \alpha }^
{0, 0, 0,\alpha} = 0,\;\;\;\;
 0 \le \alpha   \le N,
\eea
and satisfying the chirality property
\bea
W_{\alpha',\beta',\gamma',\delta'}^{\alpha \;,\beta \;,\gamma \;,\delta}(\eta) 
&=&
W_{\gamma',\delta',\alpha',\beta'}^{\gamma \;,\delta \;,\alpha \;,\beta}
(-\eta),
\eea
where $\eta$ is the asymmetry parameter. We choose the 
energy scale such that all single-particle hopping couplings have a unit value,
\bea
W_{0, \alpha, 0, 0}^{0, 0, 0, \alpha} = t = 1  \; \; \; \;0 < \alpha \le N.
\eea

Let us consider initially the case of a single particle on the otherwise 
empty chain. As a consequence of translational invariance of the Hamiltonian 
(1) the eigenfunctions are the plane waves of wavenumber $k$,
\bea
\Psi_1 &=& \sum_{x_1}f(x_1,\alpha_1)E_{x_1}^{0,\alpha_1|0,0} 
|0,\ldots,0>,\nonumber\\
f(x_1,\alpha_1) &=& e^{ikx},\;\;\;\;\; k = \frac{2\pi}{L}l \;\;\;
(l =0,1,...,L-1),
\eea
with energy
\bea
E = -2\cos k.
\eea
In (7) $|0,\ldots,0>$ is the reference state with no particles. 

The eigenfunctions of the Hamiltonian (1) in the general case, where we have 
$n$   particles, 
\bea
\Psi_n =\sum_{\{Q\}}\sum_{\{ x\}} 
f(x_1,\alpha_1;x_2,\alpha_2;...;x_n,\alpha_n)E_{x_1}^{0,\alpha_1|0,0}
E_{x_2}^{0,\alpha_2|0,0}...E_{x_n}^{0,\alpha_n|0,0} |0.\ldots,0>, 
\eea
are calculated by using a generalized nested-Bethe ansatz. In (9) the first 
summation is over all the permutations 
$Q = [Q_1, Q_2,\ldots,Q_n]$ of the 
integers $1,2,\ldots,N$, and for a given permutation $[Q]$ the second sum 
is restricted to the set $1\leq x_{Q_1} \leq x_{Q_2} \leq \cdots \leq 
x_{Q_n}$. We seek the amplitudes $f$ in each of these regions 
in the form of a superposition of plane waves with wavenumbers 
$k_j, j=1,\ldots,n$.  If we 
have only single occupation $(x_{Q_i} \neq x_{Q_{i+1}}, i = 1,2,\ldots,n-1)$ 
we write the ansatz 
\bea
f(x_1,\alpha_{Q_1};x_2,\alpha_{Q_2};...;x_n,\alpha_{Q_n}) = \sum_{P}
A_{P_1...\;\;P_n}^{\alpha_{Q_1}...\alpha_{Q_n}}
\prod_{j=1}^n\mbox{exp}(ik_{P_j}x_{Q_j}),
\eea
where the sum is over all permutations $P =[P_1,\ldots,P_n]$ of the 
integers $1,2,\ldots,n$. In the case we have a pair at the position 
$x_{Q_l} = x_{Q_{l+1}}$, the ansatz is modified to 
\bea
f(x_1,\alpha_{Q_1};x_2,\alpha_{Q_2};...;x_n,\alpha_{Q_n}) = \sum_{P}
A_{P_1...\;\;P_lP_{l+1}...P_n}^{\alpha_{Q_1}...
\overline{\alpha_{Q_l}\alpha_{Q_{l+1}}}...\alpha_{Q_n}}
\prod_{j=1}^n\mbox{exp}(ik_{P_j}x_{Q_j}),
\eea
where the bar at the $l$th and $(l+1)$th positions of the superscript 
indicates the pair location.  The general case with many isolated particles 
and pairs follows from (10) and (11). The ansatz (10)-(11) corresponding 
to the configurations where $|x_{Q_{i+1}} - x_{Q_i}| >1$ satisfies
$H|\Psi> = E |\Psi>$, providing the energy and momentum are given by 
\bea 
E = -2\sum_{j=1}^n \cos k_j; \;\;\; P = \sum_{j=1}^n k_j.
\eea 
In order to obtain the two-particle scattering matrix we consider the 
general amplitudes  (10) or (11) in the case where there are 
only two 
particles on  two neighboring sites. In this case our problem is
exactly equivalent to that of a model with only two species (bands). 
The coefficients
$A_{P_1...P_n}^{\alpha_{Q_1}...\alpha_{Q_n}}$ arising from the 
different permutation $Q$ are connected with each other by the elements
of the two-particle S-matrix
\bea
A_{...P_1 P_2...}^{...\alpha\beta ...} = 
\sum_{\delta,\gamma=1}^N S_{\alpha\beta}^{\gamma\delta}(k_{P_1},k_{P_2})
A_{...P_2 P_1...}^{...\delta \gamma...}
\eea
As a necessary condition to ensure integrability, the two-particle scattering 
matrix
has to satisfy the Yang-Baxter equation [12 - 13]. There are two integrable 
models, described in terms of two species of particles, whose S-matrix has 
a factorizable form. The 
first of these models is the standard Hubbard model [13], while the second 
one is   the correlated hopping model [14 - 16]. We may try to use as the 
fundamental building block of a general S-matrix of our model (1) 
the two-particle scattering matrix of these models. Here we consider the 
second model  and choose the S-matrix as in the correlated hopping 
model [16]
{\footnote { Unfortunately previous attempts to solve this problem by using 
the Hubbard S-matrix was unsuccessful [17] since the Bethe ansatz does not 
work in the sector where we have three or more particles. The interactions 
necessary to prevent more than double ocupancy per site spoils the exact 
integrability. This is not the case if we use as the fundamental building 
block the S-matrix of the correlated hopping model.}.
 In this case we have the following restrictions on the parameters 
of the Hamiltonian (1)
\bea
W_{0, \alpha, 0, \beta}^{0, \alpha, 0, \beta} &=& 0,\;\;\;0 <  \alpha < \beta
\nonumber\\
W_{\alpha, \beta, 0, 0}^{\alpha ,\beta, 0, 0}  &=& \epsilon_0\;\;\;\;
 0 < \alpha \le \beta \nonumber\\
W_{0, \alpha, 0, \beta}^{\alpha, \beta, 0, 0}  &=& 
W_{0, \alpha, 0, \beta}^{0, 0, \alpha, \beta } = \sqrt{1 + e^{2\eta}},\;\;\;
0 <  \alpha < \beta\nonumber\\
W_{0, \alpha, 0, \alpha}^{\alpha, \alpha, 0, 0}  &=& 2\epsilon_1\cosh\eta,\;\;\;
\alpha \in \aleph_2 \nonumber\\
W_{\alpha, \beta, 0, 0}^{0, 0, \alpha, \beta}  &=& \epsilon_0\;\;\;\;
 0 < \alpha \le \beta, 
\eea
where $\epsilon_0 = \pm 1$ and $\epsilon_1 = \pm 1$.
The non-vanishing elements of the S-matrix are 
\bea
S_{\alpha \beta}^{\alpha \beta}(k_1,k_2) &=& -\frac{\sin(\lambda_1 - \lambda_2)}
{\sin(\lambda_1 - \lambda_2 - i\eta)} \nonumber\\
S_{\beta\alpha}^{\alpha \beta}(k_1,k_2) &=& \frac{-i\sinh\eta}
{\sin(\lambda_1 - \lambda_2 - i\eta)}\mbox{exp}[i\mbox{sign}(\alpha - \beta)
(\lambda_1 - \lambda_2)] \nonumber\\
S_{\alpha \alpha}^{\alpha \alpha}(k_1,k_2) &=& -\frac{\sin[-\varepsilon_i(\lambda_1 - \lambda_2) - i\eta]}
{\sin(\lambda_1 - \lambda_2 - i\eta)} 
\eea 
where
\bea
e^{ik_j} = -\epsilon_0\frac{\sin(\lambda_j + i\eta)}{\sin(\lambda_j - i\eta)} .
\eea

To complete the proof of the Bethe ansatz (10-11) we must consider the 
eigenvalue equations in the case where there are three and four particles
on the two neighboring sites $j$ and $j+1$. This gives us a complicated system 
of equations for those  parameters of the Hamiltonian (1) involving three 
and four particles.
 We have treated this system analytically and checked our results 
numerically. We found the following solution for the diagonal elements of $W$
\bea
W_{\alpha, \beta, 0, \gamma}^{\alpha, \beta, 0, \gamma} &=& 
\epsilon_0 (1 - e^{2\eta}),\;\;\;
0 <  \alpha \le \beta < \gamma ,\nonumber\\
W_{\alpha, \beta, 0, \gamma}^{\alpha, \beta, 0, \gamma} &=& 
\epsilon_0 (1 - e^{-2\eta}),\;\;\;
0 <  \gamma < \alpha \le \beta ,\nonumber\\
W_{\alpha, \beta, 0, \gamma}^{\alpha, \beta, 0, \gamma} &=& 
0 ,\;\;\;
0 < \alpha <  \gamma < \beta ,\nonumber\\
W_{\alpha, \beta, 0, \beta}^{\alpha, \beta, 0, \beta} &=& \cases{       
\epsilon_0,& $\beta \in  \aleph_1$\cr       
-\epsilon_0 e^{2\eta},& $\beta \in \aleph_2$\cr}   ,\;\;\;
 0 <  \alpha  < \beta , \nonumber\\
W_{\alpha, \beta, 0, \alpha}^{\alpha, \beta, 0, \alpha} &=& \cases{       
\epsilon_0,& $\alpha \in  \aleph_1$\cr       
-\epsilon_0 e^{-2\eta},& $\alpha \in \aleph_2$\cr}   ,\;\;\;
 0 < \alpha  < \beta, \nonumber\\
W_{\alpha, \beta, \gamma, \delta}^{\alpha, \beta, \gamma, \delta} &=& 
\epsilon_0 (1 - 2 e^{2\eta}),\;\;\;
0 <  \alpha \le \beta < \gamma \le \delta,\nonumber\\
W_{\alpha, \beta, \gamma, \delta}^{\alpha, \beta, \gamma, \delta} &=& 
-\epsilon_0 e^{2\eta},\;\;\;
0 <  \alpha < \gamma < \beta < \delta, \nonumber\\
W_{\alpha, \beta, \gamma, \delta}^{\alpha, \beta, \gamma, \delta} &=& 
\epsilon_0 (1 - 2 \cosh 2\eta),\;\;\;
0 <  \alpha < \gamma \le \delta < \beta, \nonumber\\
W_{\alpha, \beta, \beta, \gamma}^{\alpha, \beta, \beta, \gamma} &=& 
\cases{       
\epsilon_0(1 - e^{2\eta}),& $\beta \in  \aleph_1$\cr       
-2\epsilon_0 e^{2\eta},& $\beta \in \aleph_2$\cr}   ,\;\;\;
 0 < \alpha   < \beta < \gamma, \nonumber\\
W_{\alpha, \gamma,  \beta,  \gamma}^{\alpha, \gamma, \beta,  \gamma} &=& 
\cases{       
\epsilon_0(1 - e^{2\eta}),& $\gamma \in  \aleph_1$\cr       
-2\epsilon_0 \cosh 2\eta,& $\gamma \in \aleph_2$\cr}   ,\;\;\;
 0 < \alpha  < \beta < \gamma, \nonumber\\
W_{\alpha, \beta,  \alpha, \gamma}^{\alpha, \beta,  \alpha, \gamma} &=& 
\cases{       
\epsilon_0(1 - e^{2\eta}),& $\alpha \in  \aleph_1$\cr       
-2\epsilon_0 \cosh 2\eta,& $\alpha \in \aleph_2$\cr}   ,\;\;\;
 0 < \alpha  < \beta < \gamma, \nonumber\\
W_{\alpha, \beta,  \alpha, \beta}^{\alpha, \beta,  \alpha, \beta} &=& 
\cases{       
2\epsilon_0 ,& $\alpha, \beta \in  \aleph_1$\cr       
-2\epsilon_0 \cosh 2\eta,& $\alpha \;\;\mbox{or}\;\;
\beta \in \aleph_2$\cr}   ,\;\;\;
 0 < \alpha  < \beta, \nonumber\\
W_{\alpha, \alpha, \alpha, \beta}^ {\alpha, \alpha, \alpha, \beta}&=& 
-\epsilon_0 (2e^{2\eta} + e^{-2\eta}),\;\;\;\alpha \in \aleph_2, 
0 <  \alpha  < \beta,\nonumber\\
W_{\beta, \beta, \alpha, \beta}^{\beta, \beta, \alpha, \beta}&=& 
-\epsilon_0 (e^{2\eta} + 2e^{-2\eta}),\;\;\;\beta \in \aleph_2, 
0 <  \alpha  < \beta,\nonumber\\
W_{\alpha, \alpha, 0, \alpha}^{\alpha, \alpha, 0, \alpha} &=& 
\frac{1}{2}W_{\alpha, \alpha, \alpha, \alpha}^{\alpha, \alpha , \alpha, \alpha} = 
-2\epsilon_0\cosh 2\eta , \;\;\; \alpha \in \aleph_2 .
\eea 
For the nondiagonal elements of $W$ we have
\bea
W_{\alpha, \beta, 0 ,\gamma}^{0 ,\alpha, \beta, \gamma} &=& 1\;\;\;\;
0 < \alpha < \beta < \gamma ,\nonumber\\
W_{\alpha, \beta, 0 ,\gamma}^{0, \beta, \alpha, \gamma} &=& 
W_{0, \alpha, \beta, \gamma}^{\alpha, \gamma, 0, \beta} = e^{-\eta}, \;\;\;\;
0 < \alpha < \beta < \gamma ,\nonumber\\
W_{\alpha, \beta,  0, \beta}^{0, \beta,  \alpha, \beta} &=& \cases{       
2\cosh\eta ,& $\beta \in  \aleph_1$\cr       
0,& $\beta \in \aleph_2$\cr}   ,\;\;\;
 0 < \alpha  < \beta, \nonumber\\
W_{\alpha, \beta,  0, \alpha}^{0, \alpha, \alpha, \beta} &=& \cases{       
2\cosh\eta ,& $\alpha \in  \aleph_1$\cr       
0,& $\alpha \in \aleph_2$\cr}   ,\;\;\;
 0 < \alpha  < \beta, \nonumber\\
W_{\alpha, \alpha, 0, \beta }^{0, \alpha, \alpha, \beta} &=& 
W_{0, \alpha, \beta, \beta}^{\alpha, \beta, 0, \beta} = \epsilon_1\sqrt{1 +
e^{-2\eta}}, \;\;\;\;
0 < \alpha < \beta ,\nonumber\\
W_{\alpha, \beta, 0, \gamma}^{\alpha, \gamma, 0, \beta} &=& 
W_{0, \alpha, \beta, \gamma}^{ 0, \beta, \alpha, \gamma } = 
-\epsilon_0 e^{\eta}, \;\;\;\;
0 < \alpha < \beta < \gamma ,\nonumber\\
W_{\alpha, \beta, 0 ,\gamma}^{\beta, \gamma, 0, \alpha } &=& 
-\epsilon_0, \;\;\;\;
0 < \alpha < \beta < \gamma ,\nonumber\\
W_{\alpha, \alpha, 0, \beta }^{\alpha, \beta, 0, \alpha} &=& 
W_{0, \alpha, \beta, \beta}^{0, \beta, \alpha, \beta} = -\epsilon_0\epsilon_1
\sqrt{1 +e^{2\eta}}, \;\;\;\;
0 < \alpha < \beta ,\nonumber\\
W_{\alpha, \beta, 0, \gamma}^{0, \gamma, \alpha, \beta } &=& 
- 1, \;\;\;\;
0 < \alpha \le \beta , \;\;\;\gamma \ne \alpha, \beta, \nonumber\\
W_{\alpha, \beta, \gamma, \delta}^{\alpha, \delta, \beta, \gamma} &=& 
W_{\alpha \beta \gamma \delta}^{\beta \gamma \alpha \delta} = -\epsilon_0, \;\;\;\;
0 < \alpha < \beta < \gamma < \delta,\nonumber\\
W_{\alpha, \beta, \gamma, \delta}^{\alpha, \gamma, \beta, \delta} &=& 
W_{\alpha, \gamma, \beta, \delta}^{\gamma, \delta, \alpha, \beta}  \nonumber\\
&=& W_{\beta, \delta, \alpha, \gamma}^{\alpha, \delta, \beta, \gamma}  
W_{\beta, \delta, \alpha, \gamma}^{\beta, \gamma, \alpha, \delta} = 
-\epsilon_0 e^{\eta}, \;\;\;\;
0 < \alpha < \beta < \gamma < \delta,\nonumber\\
W_{\alpha, \alpha, \beta, \gamma}^{\alpha, \beta, \alpha, \gamma} &=& 
W_{\alpha, \beta, \gamma, \gamma}^{\alpha, \gamma, \beta, \gamma} = 
W_{\beta, \gamma, \alpha, \alpha}^{\alpha, \beta, \alpha, \gamma} = 
W_{\gamma, \gamma, \alpha, \beta }^{\alpha, \gamma, \beta, \gamma } = 
W_{\beta, \beta, \alpha, \gamma}^{\beta, \gamma, \alpha, \beta} = 
W_{\alpha, \gamma, \beta, \beta}^{\beta, \gamma, \alpha, \beta}  \nonumber\\
&=&W_{\alpha, \gamma, \beta, \beta}^{\beta, \gamma, \alpha, \beta}  
W_{\beta, \beta, \alpha, \gamma}^{\beta, \gamma, \alpha, \beta} = 
-\epsilon_0\epsilon_1
\sqrt{1 +e^{2\eta}}, \;\;\;\;
0 < \alpha < \beta < \gamma, \nonumber\\
W_{\alpha, \beta, \beta, \gamma}^{\beta, \gamma, \alpha, \beta} &=& 
W_{\alpha, \beta, \gamma, \beta}^{\gamma, \beta, \alpha, \beta} = 
W_{\beta, \alpha, \beta, \gamma}^{\beta, \gamma, \beta, \alpha} =  
\cases{       
-2\epsilon_0\cosh\eta ,& $\beta \in  \aleph_1$\cr       
0,& $\beta \in \aleph_2$\cr}   ,\;\;\;
 0 < \alpha  \ne \beta \ne \gamma, \nonumber\\
W_{\alpha, \beta, \alpha, \alpha}^{\alpha, \alpha, \alpha, \beta} &=& 
W_{\alpha, \beta, \beta, \beta }^{\beta, \beta, \alpha, \beta} = 
-\epsilon_0, \;\;\;\;
0 < \alpha < \beta,\nonumber\\
W_{\alpha, \beta, \alpha, \beta}^{\alpha, \alpha, \beta, \beta} &=& 
-2\epsilon_0\cosh\eta\;\;\;
0 < \alpha < \beta,\nonumber\\
W_{\alpha,  \beta, \gamma, \delta}^{\gamma, \delta, \alpha, \beta} &=&
\epsilon_0, \;\;\;\;
\alpha , \beta \ne \gamma, \delta.
\eea

The periodic boundary condition for the system on the finite lattice, with 
  size
$L$, gives us the Bethe-ansatz equations. In order to obtain these equations 
we must diagonalize the transfer matrix of a related inhomogeneous vertex 
model with non-intersecting strings [18]. The Bethe-ansatz equations are 
written in terms of the momenta of the electrons $k_j$ and additional rapidities
$\Lambda_{\alpha}^{i}$
\bea
\lbrack\frac{\sin(\lambda_j + i\eta)}{\sin(\lambda_j - i\eta)}\rbrack^L &=&
\varepsilon_1^{n-m_1-1}\prod_{j'=1}^n\frac{\sin(\lambda_j - \lambda_{j'} + 
i\varepsilon_1\eta)}{\sin(\lambda_j - \lambda_{j'} - i\eta)}
\prod_{\alpha=1}^{m_1}\frac{\sin(\lambda_j-\Lambda_{\alpha}^{(1)} - \frac{i}{2}
\varepsilon_1\eta)}{\sin(\lambda_j-\Lambda_{\alpha}^{(1)} + \frac{i}{2}
\varepsilon_1\eta)}
\nonumber\\
\prod_{\alpha'=1}^{m_{\sigma}}\frac{\sin(\Lambda_{\alpha}^{(\sigma)} - \Lambda_{\alpha'}^{(\sigma)} +i\varepsilon_{\sigma+1}\eta)}{\sin(\Lambda_{\alpha}^{(\sigma)} - \Lambda_{\alpha'}^{(\sigma)} -i\varepsilon_{\sigma}\eta)} &=&
\varepsilon_{\sigma}^{m_{\sigma-1}+m_{\sigma}-1}
\varepsilon_{\sigma+1}^{m_{\sigma+1}+m_{\sigma}-1}
\prod_{\alpha'=1}^{m_{\sigma-1}}\frac{\sin(\Lambda_{\alpha}^{(\sigma)} -
\Lambda_{\alpha}^{(\sigma-1)} +\frac{i}{2}\varepsilon_{\sigma}\eta)}
{\sin(\Lambda_{\alpha}^{(\sigma)} -
\Lambda_{\alpha}^{(\sigma-1)} -\frac{i}{2}\varepsilon_{\sigma}\eta)}\nonumber\\
&\times&\prod_{\alpha'=1}^{m_{\sigma+1}}\frac{\sin(\Lambda_{\alpha}^{(\sigma)} -
\Lambda_{\alpha}^{(\sigma+1)} +\frac{i}{2}\varepsilon_{\sigma+1}\eta)}
{\sin(\Lambda_{\alpha}^{(\sigma)} -
\Lambda_{\alpha}^{(\sigma+1)} -\frac{i}{2}\varepsilon_{\sigma+1}\eta)},
\eea
\bea
\sigma = 1,2,...,N-1,\;\;\; m_0 = n,\;\;\; m_N = 0, \;\;\;
\Lambda_{\alpha}^{(0)} = \lambda_{\alpha},\nonumber
\eea
where $n_j = m_{j-1} - m_j$ is the number of particles of species $j$, 
 $\varepsilon_{\sigma} = +1$ if $\sigma \in \aleph_1$ and 
$\varepsilon_{\sigma}=-1$ for $\sigma  \in 
\aleph_2$.
The energy of the system is given in terms of the Bethe-ansatz roots 
$\lambda_j$ 
\bea
E = -2\sum_{j=1}^n \cos k_j = 2\epsilon_0\sum_{j=1}^n\lbrack\cosh2\eta -
\frac{\sinh^22\eta}{\cosh 2\eta -\cos 2\lambda_j}\rbrack .
\eea
It is interesting to remark that the 
 particular isotropic cases  $\gamma = 0$, of (19 - 20), where $N=2$ and  
$\varepsilon_1 = - \varepsilon_2 = 1$, as well as for arbitrary $N$, but 
  $\varepsilon_i = 1 (i=1, \ldots, N)$ 
 have been obtained in refs. [19] and [20], respectively.

In conclusion, we have studied a new integrable model which can be presented 
as a system of interacting XX and Fateev-Zamolodchikov chains. 
The Bethe-ansatz 
 equations are derived by means of the coordinate Bethe-ansatz
approach. A  desirable continuation of the present work is the investigation 
of the  thermodynamic
equilibrium properties of the model based on the Bethe-ansatz solutions  
(19 - 20).

This work was supported in part by Conselho Nacional        
de Desenvolvimento 
Cient\'{\i}fico e Tecnol\'ogico  - CNPq - Brazil, by FINEP - Brazil and by the 
Russian Foundation of Fundamental Investigation (Grant 99-02017646).

\end{document}